\documentclass[a4paper,12pt]{article}

\usepackage{latexsym}

\newcommand{\nc}{\newcommand}

\nc{\be}[1]{\begin{equation}\mbox{$\label{#1}$}}
\nc{\bea}[1]{\begin{eqnarray} \mbox{$\label{#1}$}}
\nc{\Section}[2]{\section{#2}\label{#1}}
\nc{\Bibitem}[1]{\bibitem{#1}}
\nc{\Label}[1]{\label{#1}}

\nc{\eea}{\end{eqnarray}}
\nc{\ee}{\end{equation}}

\nc{\bdm}{\begin{displaymath}}
\nc{\edm}{\end{displaymath}}
\nc{\dpsty}{\displaystyle}
\nc{\bc}{\begin{center}}
\nc{\ec}{\end{center}}
\nc{\ba}{\begin{array}}
\nc{\ea}{\end{array}}
\nc{\bab}{\begin{abstract}}
\nc{\eab}{\end{abstract}}
\nc{\btab}{\begin{tabular}}
\nc{\etab}{\end{tabular}}
\nc{\bit}{\begin{itemize}}
\nc{\eit}{\end{itemize}}
\nc{\ben}{\begin{enumerate}}
\nc{\een}{\end{enumerate}}
\nc{\bfig}{\begin{figure}}
\nc{\efig}{\end{figure}}

\nc{\arreq}{&\!=\!&}
\nc{\arrmi}{&\!-\!&}
\nc{\arrpl}{&\!+\!&}
\nc{\arrap}{&\!\!\!\approx\!\!\!&}
\nc{\non}{\nonumber}
\nc{\align}{\!\!\!\!\!\!\!\!&&}

\def\lsim{\; \raise0.3ex\hbox{$<$\kern-0.75em
      \raise-1.1ex\hbox{$\sim$}}\; }
\def\gsim{\; \raise0.3ex\hbox{$>$\kern-0.75em
      \raise-1.1ex\hbox{$\sim$}}\; }

\nc{\DOT}{\hspace{-0.08in}{\bf .}\hspace{0.1in}}
\nc{\Laada}{\hbox {$\sqcap$ \kern -1em $\sqcup$}}
\nc\loota{{\scriptstyle\sqcap\kern-0.55em\hbox{$\scriptstyle\sqcup$}}}
\nc\Loota{{\sqcap\kern-0.65em\hbox{$\sqcup$}}}
\nc\laada{\Loota}
\nc{\qed}{\hskip 3em \hbox{\BOX} \vskip 2ex}

\nc{\real}{{\rm I \! R}}
\nc{\Z}{{\sf Z \!\!\! Z}}
\nc{\complex}{{\rm C\!\!\! {\sf I}\,\,}}
\def\bigid{\leavevmode\hbox{\small1\kern-3.8pt\normalsize1}}
\def\id{\leavevmode\hbox{\small1\kern-3.3pt\normalsize1}}
\nc{\slask}{\!\!\!/}
\nc{\bis}{{\prime\prime}}
\nc{\pa}{\partial}
\nc{\na}{\nabla}
\nc{\ra}{\rangle}
\nc{\la}{\langle}
\nc{\goto}{\rightarrow}
\nc{\swap}{\leftrightarrow}

\nc{\EE}[1]{ \mbox{$\cdot10^{#1}$} }
\nc{\abs}[1]{\left|#1\right|}
\nc{\at}[2]{\left.#1\right|_{#2}}
\nc{\norm}[1]{\|#1\|}
\nc{\abscut}[2]{\Abs{#1}_{\scriptscriptstyle#2}}
\nc{\vek}[1]{{\rm\bf #1}}
\nc{\integral}[2]{\int\limits_{#1}^{#2}}
\nc{\inv}[1]{\frac{1}{#1}}
\nc{\dd}[2]{{{\partial #1}\over{\partial #2}}}
\nc{\ddd}[2]{{{{\partial}^2 #1}\over{\partial {#2}^2}}}
\nc{\dddd}[3]{{{{\partial}^2 #1}\over
    {\partial #2 \partial #3}}}
\nc{\dder}[2]{{{d #1}\over{d #2}}}
\nc{\ddder}[2]{{{d^2 #1}\over{d {#2}^2}}}
\nc{\dddder}[3]{{d^2 #1}\over
    {d #2 d #3}}
\nc{\dx}[1]{d\,^{#1}x}
\nc{\dy}[1]{d\,^{#1}y}
\nc{\dz}[1]{d\,^{#1}z}
\nc{\dl}[1]{\frac{d\,^{#1}l}{(2\pi)^{#1}}}
\nc{\dk}[1]{\frac{d\,^{#1}k}{(2\pi)^{#1}}}
\nc{\dq}[1]{\frac{d\,^{#1}q}{(2\pi)^{#1}}}

\nc{\bfT}{{\bf T }}
\def\GeV{{\rm\ GeV}}
\def\MeV{{\rm MeV}}
\def\keV{{\rm\ keV}}
\def\TeV{{\rm\ TeV}}
\def\cm{{\rm\ cm}}
\def\g{{\rm\ g}}

\nc{\cA}{{\cal A}}
\nc{\cB}{{\cal B}}
\nc{\cD}{{\cal D}}
\nc{\cE}{{\cal E}}
\nc{\cG}{{\cal G}}
\nc{\cH}{{\cal H}}
\nc{\cL}{{\cal L}}
\nc{\cO}{{\cal O}}
\nc{\cT}{{\cal T}}
\nc{\cN}{{\cal N}}
\nc{\rvac}[1]{|{\cal O}#1\rangle}
\nc{\lvac}[1]{\langle{\cal O}#1|}
\nc{\rvacb}[1]{|{\cal O}_\beta #1\rangle}
\nc{\lvacb}[1]{\langle{\cal O}_\beta #1 |}
\nc{\bb}{\bar{\beta}}
\nc{\bt}{\tilde{\beta}}
\nc{\ctH}{\tilde{\cal H}}
\nc{\chH}{\hat{\cal H}}
\nc{\1}{\aa}
\nc{\2}{\"{a}}
\nc{\3}{\"{o}}
\nc{\4}{\AA}
\nc{\5}{\"{A}}
\nc{\6}{\"{O}}
\nc{\al}{\alpha}

\nc{\Del}{\Delta}
\nc{\e}{\textrm{e}}
\nc{\eps}{\epsilon}
\nc{\lam}{\lambda}
\nc{\Om}{\Omega}
\nc{\ve}{\varepsilon}
\nc{\mn}{{\mu\nu}}
\nc{\vp}{\varphi}
\nc{\vev}[1]{\langle #1 \rangle}
\nc{\advp}[3]{{\it  Adv.\ in\ Phys.\ }{{\bf #1} {(#2)} {#3}}}
\nc{\annp}[3]{{\it  Ann.\ Phys.\ (N.Y.)\ }{{\bf #1} {(#2)} {#3}}}
\nc{\apl}[3]{{\it  Appl. Phys. Lett. }{{\bf #1} {(#2)} {#3}}}
\nc{\apj}[3]{{\it  Ap.\ J.\ }{{\bf #1} {(#2)} {#3}}}
\nc{\apjl}[3]{{\it  Ap.\ J.\ Lett.\ }{{\bf #1} {(#2)} {#3}}}
\nc{\app}[3]{{\it Astropart.\ Phys.\ }{{\bf #1} {(#2)} {#3}}}
\nc{\cmp}[3]{{\it  Comm.\ Math.\ Phys.\ }{{ \bf #1} {(#2)} {#3}}}
\nc{\cqg}[3]{{\it  Class.\ Quant.\ Grav.\ }{{\bf #1} {(#2)} {#3}}}
\nc{\epl}[3]{{\it  Europhys.\ Lett.\ }{{\bf #1} {(#2)} {#3}}}
\nc{\ijmp}[3]{{\it Int.\ J.\ Mod.\ Phys.\ }{{\bf #1} {(#2)} {#3}}}
\nc{\ijtp}[3]{{\it Int.\ J.\ Theor.\ Phys.\ }{{\bf #1} {(#2)} {#3}}}
\nc{\jmp}[3]{{\it  J.\ Math.\ Phys.\ }{{ \bf #1} {(#2)} {#3}}}
\nc{\jpa}[3]{{\it  J.\ Phys.\ A\ }{{\bf #1} {(#2)} {#3}}}
\nc{\jpc}[3]{{\it  J.\ Phys.\ C\ }{{\bf #1} {(#2)} {#3}}}
\nc{\jap}[3]{{\it J.\ Appl.\ Phys.\ }{{\bf #1} {(#2)} {#3}}}
\nc{\jpsj}[3]{{\it J.\ Phys.\ Soc.\ Japan\ }{{\bf #1} {(#2)} {#3}}}
\nc{\lmp}[3]{{\it Lett.\ Math.\ Phys.\ }{{\bf #1} {(#2)} {#3}}}
\nc{\mpl}[3]{{\it  Mod.\ Phys.\ Lett.\ }{{\bf #1} {(#2)} {#3}}}
\nc{\ncim}[3]{{\it  Nuov.\ Cim.\ }{{\bf #1} {(#2)} {#3}}}
\nc{\np}[3]{{\it  Nucl.\ Phys.\ }{{\bf #1} {(#2)} {#3}}}
\nc{\pr}[3]{{\it Phys.\ Rev.\ }{{\bf #1} {(#2)} {#3}}}
\nc{\pra}[3]{{\it  Phys.\ Rev.\ A\ }{{\bf #1} {(#2)} {#3}}}
\nc{\prb}[3]{{\it  Phys.\ Rev.\ B\ }{{{\bf #1} {(#2)} {#3}}}}
\nc{\prc}[3]{{\it  Phys.\ Rev.\ C\ }{{\bf #1} {(#2)} {#3}}}
\nc{\prd}[3]{{\it  Phys.\ Rev.\ D\ }{{\bf #1} {(#2)} {#3}}}
\nc{\prl}[3]{{\it Phys\ Rev.\ Lett.\ }{{\bf #1} {(#2)} {#3}}}
\nc{\pl}[3]{{\it  Phys.\ Lett.\ }{{\bf #1} {(#2)} {#3}}}
\nc{\prep}[3]{{\it Phys\. Rep.\ }{{\bf #1} {(#2)} {#3}}}
\nc{\prsl}[3]{{\it Proc.\ R.\ Soc.\ London\ }{{\bf #1} {(#2)} {#3}}}
\nc{\ptp}[3]{{\it  Prog.\ Theor.\ Phys.\ }{{\bf #1} {(#2)} {#3}}}
\nc{\ptps}[3]{{\it  Prog\ Theor.\ Phys.\ suppl.\ }{{\bf #1} {(#2)} {#3}}}
\nc{\physa}[3]{{\it  Physica\ A\ }{{\bf #1} {(#2)} {#3}}}
\nc{\physb}[3]{{\it  Physica\ B\ }{{\bf #1} {(#2)} {#3}}}
\nc{\phys}[3]{{\it Physica\ }{{\bf #1} {(#2)} {#3}}}
\nc{\rmp}[3]{{\it  Rev.\ Mod.\ Phys.\ }{{\bf #1} {(#2)} {#3}}}
\nc{\rpp}[3]{{\it Rep.\ Prog.\ Phys.\ }{{\bf #1} {(#2)} {#3}}}
\nc{\sjnp}[3]{{\it Sov.\ J.\ Nucl.\ Phys.\ }{{\bf #1} {(#2)} {#3}}}
\nc{\spjetp}[3]{{\it Sov.\ Phys.\ JETP\ }{{\bf #1} {(#2)} {#3}}}
\nc{\yf}[3]{{\it Yad.\ Fiz.\ }{{\bf #1} {(#2)} {#3}}}
\nc{\zetp}[3]{{\it Zh.\ Eksp.\ Teor.\ Fiz.\  }{{\bf #1}  {(#2)} {#3}}}
\nc{\zp}[3]{{\it Z.\ Phys.\ }{{\bf #1} {(#2)} {#3}}}
\nc{\ibid}[3]{{\sl ibid.\ }{{\bf #1} {#2} {#3}}}

\nc{\rf}[1]{(\ref{#1})}
\nc{\nn}{\nonumber \\*}
\nc{\bfB}{\bf{B}}
\nc{\bfv}{\bf{v}}
\nc{\bfx}{\bf{x}}
\nc{\bfy}{\bf{y}}
\nc{\vx}{\vec{x}}
\nc{\vy}{\vec{y}}
\nc{\oB}{\overline{B}}
\nc{\oI}{\overline{I}}
\nc{\oR}{\overline{R}}
\nc{\rar}{\rightarrow}
\nc{\ti}{\times}
\nc{\slsh}{\hskip-5pt/}
\nc{\sm}{Standard~Model~}
\nc{\MP}{M_{\rm Pl}}
\nc{\tp}{t_{\rm Pl}}

\nc{\pmin}{p_{\rm min}}
\nc{\pmax}{p_{\rm max}}
\nc{\fo}{f_0}
\nc{\foi}{f_{0,i}\,}
\nc{\fop}{f_0^P}
\nc{\fou}{f_0^U}

\nc{\eff}{{\rm eff}}
\nc{\MT}{M_{\rm T}}
\nc{\ML}{M_{\rm L}}
\nc{\kk}{\vek{k}}
\nc{\pp}{{\rm p}}
\nc{\pt}{\partial_t}
\nc{\half}{{1\over 2}}
\nc{\w}{\omega}
\nc{\uhat}{\hat{U}_\w}

\nc{\etal}{\mbox{\it et al.}}
\nc{\ie}{{\it i.e.\ }}
\nc{\eg}{{\it e.g.\ }}
\nc{\trh}{T_{\rm RH}}
\nc{\iso}[2]{#1\times 10^{#2}}
\nc{\ind}[1]{\textrm{\scriptsize #1}}
\nc{\note}[1]{\begin{flushright}\framebox{\emph{#1}}\end{flushright}}

\begin{document}
{\title{\vskip-2truecm{\hfill {{\small HIP-2001-62/TH\\
  \hfill {\small TURKU-FL-P38-01}  \hfill \\
    }}\vskip 1truecm}
{\LARGE Constraints on Self-Interacting Q-ball Dark Matter}
\vspace{-.2cm}}}

\author{
{\sc Kari Enqvist\footnote{email: kari.enqvist@helsinki.fi}}\\
{\sl\small Physics Department, University of Helsinki, and Helsinki Institute
of Physics}\\
{\sl\small P.O. Box 9, FIN-00014 University of Helsinki, FINLAND}\\
{\sc Asko Jokinen\footnote{email: asko.jokinen@helsinki.fi}}\\
{\sl\small Physics Department, University of Helsinki}\\
{\sl\small P.O. Box 9, FIN-00014 University of Helsinki, FINLAND}\\
{\sc Tuomas Multam\"aki\footnote{email: tuomul@utu.fi} and
Iiro Vilja\footnote{email: vilja@utu.fi}}\\
{\sl\small Department of Physics, University of Turku, FIN-20014, FINLAND}}

\date{November 27, 2001}

\maketitle

\begin{abstract}
We consider different types of Q-balls as self-interacting dark matter.
For the Q-balls to act as the dark matter of the universe
they should not evaporate, which requires them to carry
very large charges; depending on the type, the minimum charge could be
as high as $Q\sim 10^{33}$ or the Q-ball coupling to ordinary matter as small
as $\sim 10^{-35}$. The cross-section-to-mass ratio needed for
self-interacting dark matter implies a mass scale of $m\sim {\cal O}(1)$ MeV
for the quanta that the Q-balls consist of, which is very difficult to achieve
in the MSSM. \end{abstract}
\vfill
\thispagestyle{empty}
\newpage
\section{Introduction}
\setcounter{page}{1}

Dark matter is widely expected to form a significant portion of
the total energy density of the universe, regardless of the lack
of direct experimental evidence. Indirect experimental evidence,
on the other hand, for the existence of dark matter is well
established: galactic rotation curves, dynamics of galaxy clusters,
and large scale flows all indicate that large amounts of dark
matter must be present on galactic halo scales.
In addition, the present cosmic microwave background (CMB)
observations strongly support a cosmological model with a significant
dark matter fraction \cite{boom}.

The cosmological model that is in good agreement with the CMB observations
and large scale structure, is based on collisionless cold dark matter.
A number discrepancies, however, between observations and numerical
simulations have been noted on galactic and sub-galactic scales.
The halo density profiles, and the number density of satellite galaxies
are examples of where the collisionless cold dark matter models are in
disagreement with observations \cite{cdmprob,spergel}.

The discrepancies between expected and observed distribution of dark matter
can be alleviated, and possibly resolved, by allowing interactions
between the dark matter constituents \cite{spergel}. The scattering of dark matter
particles in high density regions leads to smoothing out of the
density distribution, randomises the velocity distribution of the dark
matter particles, and can lead to enhanced destruction of halo substructure.
All of these processes help to resolve the problems associated
with collisionless cold dark matter models.

To have the appropriate properties, the self-interaction cross-section
and mass of the dark matter particles undergoing elastic scattering
need to satisfy \cite{spergel,kustein},
\be{srange}
s={\sigma_{DD}\over m_{DM}}
\simeq  2 \times 10^3\ -\ 3 \times 10^4 \GeV^{-3}
= 0.5-6\cm^2\g^{-1},
\ee
where $\sigma_{DD}$ is the self-interaction cross-section, and
$m_{DM}$ the mass of the dark matter particle. Hence the required
interaction is relatively strong and one talks about strongly
interacting dark matter (SIDM).

Note that here only elastic scattering are considered and if other
types of processes are studied, these values may be somewhat different.
As an example, in \cite{craig},
an effective annihilation cross section per unit mass is found
to be $0.03\ {\rm cm}^2/{\rm g}$ in a model where dark matter undergoes both
elastic scattering and annihilation.

Q-balls \cite{cole} have been recently proposed as a candidate
for the self-interacting dark matter \cite{kustein}.
Q-balls are non-topological solitons \cite{leepang} that can exist
in theories with scalar fields carrying a conserved $U(1)$-charge
\cite{cole}. The Q-ball is the ground state of the theory in the
sector of fixed charge, \ie the energy of the Q-ball configuration
is less than that of a collection free scalars carrying an equal
amount charge as the Q-ball. The phase of the Q-ball field, $\phi$,
rotates uniformly with frequency $\w$, and we can write
\be{ballfield}
\phi=\varphi \e^{i\w t},
\ee
where $\varphi$ is a spherically symmetric, monotonically decreasing, positive
function \cite{cole}. The energy and charge of a Q-ball are given by
\bea{eandq}
E & = & \int dx^3[\w^2\varphi^2+|\nabla\varphi|^2+U(\varphi^2)],\\
Q & = & 2\w \int dx^3\varphi^2,
\eea
where $U(|\phi|^2)$ is the potential that has a global minimum at the origin
and is invariant under $U(1)$-transformations of the $\phi$-field.
For the Q-ball to be energetically stable, condition
\be{econd}
E<mQ,
\ee
where $m$ is the mass of the free $\phi$ scalar, must hold. This condition
is met whenever the potential $U(\varphi^2)$ grows slower than $\varphi^2$.

Stable Q-balls exist in many theories, and in particular
in supersymmetric extensions of the Standard Model, where
Q-balls are made of squarks and sleptons.

The purpose of the present
paper is to study the general circumstances under which Q-balls can act as
strongly interacting dark matter. The key point here is that although the
Q-balls typically consist of weakly interacting quanta, they are large
objects and can have large interaction rates.
This was first pointed out in a recent paper by Kusenko and Steinhardt
\cite{kustein}, where the idea of self-interacting Q-ball dark matter was
proposed. In this paper we study the proposal in more detail, taking into
account the commonly considered Q-ball types and the implications of a
primordial Q-ball charge distribution. We also discuss the importance of
evaporation and thermal processes, which need to be accounted for in any realistic
detailed model. In Section 2 we recall the salient features
of the three main types of Q-balls: thin-wall, thick-wall in flat potentials,
and thick-wall in logarithmic potentials. In Section 3 we study experimental
constraints on a dark matter Q-balls by studying their interaction cross 
sections and number distribution. In Section 4 we discuss 
how Q-ball evolution, in particular evaporation of charge from Q-balls 
and the surrounding thermal bath,
constrain the properties of Q-ball dark matter. Section 5 contains our
conclusions.

\section{Different Q-ball Types}

\subsection{Type I: Thin-wall Q-balls}
The Q-balls considered in the literature have very different properties, which
depend on the details of the potential. These differences are also reflected
in their scattering cross sections and hence in dark matter properties.
The Q-balls may either have a narrow, well-defined edge, in which case
they are called thin-wall Q-balls, or their  boundaries are not
localised in a narrow region,  in which case they are called thick-wall
Q-balls. Both types may exist within a same theory.

Let us first consider thin-wall Q-balls, which arise for
any suitable potential that allows Q-balls to exist and grows faster than
$\varphi^2$ in the large $\varphi$ limit.
These solutions are approximated by the profile $\varphi(r)\approx\varphi_0\theta(r-R)$.
The energy to charge ratio of such a configuration
is (neglecting all surface terms)
\be{tweq}
{E\over Q}=\textrm{min}\sqrt{U(\varphi^2)\over\varphi^2}\equiv\w_c,
\ee
\ie energy grows linearly with charge. Note that the radius of such a Q-ball
can be very large and is related to the charge simply by
\be{twcharge}
Q=2\w_c\varphi_0^2V={8\over 3}\pi R^3\w_c\varphi_0^2.
\ee

\subsection{Type II: Thick-wall Q-balls in flat potentials}
In addition to the thin-wall Q-balls, two other types of Q-balls have been
commonly considered. These arise {\it e.g.} in
supersymmetric theories with gauge and gravity mediated supersymmetry breaking
scenarios.

In flat potentials the mass of a Q-ball can grow more slowly that in the thin-wall case.
If the potential has an absolutely flat plateau at large $\varphi$,
$U(\varphi)\sim m^4$,
as has been studied in association with the gauge mediated supersymmetry breaking,
the energy of such a Q-ball grows as \cite{dvali}
\be{gaugeen}
E\approx {4\sqrt{2}\over 3}m\pi Q^{3/4}.
\ee
The radius and value of the field inside the Q-ball are given by
\bea{gaugerphi}
R & \approx & {1\over\sqrt{2}m}Q^{1/4}\\
\varphi_0 & \approx & {m\over\sqrt{2\pi}}Q^{1/4}.
\eea

\subsection{Type III: Thick-wall Q-balls in logarithmic potentials}
The potential may also grow only slightly slower that bare $\varphi^2$ -term.
{\it E.g.} in the gravity mediated supersymmetry breaking scenario the scalar
potential grows like \be{gravpot} U(\varphi)=m^2(1+K\log({\varphi^2\over
M^2}))\varphi^2, \ee
where $K<0$ and $M$ is a large mass scale. In potentials of this form
the Q-ball profile can be solved:
\be{gravsolu}
\varphi(r)=Me^{-(1-{\omega^2\over m^2}-2K)/(2K)}e^{Km^2r^2/2}\equiv
\tilde{M}e^{Km^2r^2/2}.
\ee
The energy charge relation is approximately
\be{graveq}
E\approx mQ.
\ee
The radius of the Q-ball remains constant with increasing charge in potentials
of this form, $R\equiv |K|^{-1/2}m^{-1}$.

Note that in the thick-wall cases where the potential grows more slowly than a
mass term, $m^2\varphi^2$, the non-renormalizable terms will begin to dominate
at some large value of $\varphi$ after which the Q-ball solution approaches the
thin-wall type. Since the potentials associated with the Type II and Type III
Q-balls represent the extremes (strong binding and weak binding, respectively),
any thick-wall Q-ball should fall into a category that is somewhere in between.
Hence it is sufficient to consider only the above three types separately.

\section{Flux limits on Q-ball dark matter}
\subsection{Scattering Cross-sections}
The scattering cross-section of Q-balls has been studied
for different types of Q-balls in \cite{kawa2}-\cite{multa2}.
Collisions of thick-wall Q-balls associated with potentials in the
gauge and gravity mediated SUSY breaking scenarios have been studied
in detail in \cite{multa1,multa2}.
There it was found that for like Q-balls, the average fusion
and charge transfer cross-sections are somewhat larger than the geometric
cross-section in the studied charge range. It was also found that
the type of the collision process: merger, elastic scattering, or charge transfer
between the Q-balls, is dependent on the relative phase of the colliding Q-balls.
If the Q-balls are in phase when they collide, they will typically fuse into one
large Q-ball where as if their phase difference is $\pi$, they will repel each other.
All of these properties were noted at velocities of the order of
typical galactic velocities. At higher velocities, the scattering cross-section
typically decreases due to a reduced interaction time as was noted in \cite{multa2}.

In the study of collision processes, it was also noted that when two equally sized
Q-balls collide, most of the charge can be transferred to one of the Q-balls,
so that one of the Q-balls in the final state is small compared to its initial
charge. In such a process the small Q-ball can acquire a large velocity due to
momentum and energy conservation. In the simulations a ten-fold increase in velocity
was not uncommon and hence such processes can reduce the number of dark matter Q-balls
in the galaxy as the small Q-ball can escape from the galactic halo.

On the basis of the numerical simulations \cite{multa1,multa2} we may write
$\sigma_{DD}=\xi \pi R^2$, where $\xi$ represents the
deviation of the scattering cross-section from the geometric one.
Typically one finds that $\xi\simeq 2$ for the gauge mediated and
$\xi\simeq 4$ for the gravity mediated case. We take $\xi=1$ for the
thin-wall case. The scattering cross-section to mass
ratio, $s = \sigma_{DD}/E_Q$, of the different types of Q-balls are given by:
\bea{diffs}
s_\ind{I} & = & {\xi\pi R^2\over \w_c Q}=({9\pi\over 64
\w_c^5\varphi_0^4})^{1/3}Q^{-1/3}~~{\rm (thin-wall)}\non\\
s_\ind{II} & \simeq &{3\xi\over 4\sqrt{2}m^3}Q^{-1/4}~~{\rm
(thick-wall,~flat~}U)\non\\
s_\ind{III} & \simeq & {4\pi\xi\over |K|m^3}Q^{-1}~~{\rm
(thick-wall,~log~}U).
\eea

Inserting the allowed values for $s$ from (\ref{srange}), we obtain
the range of acceptable charges:
\bea{qranges}
\textrm{I}: & 3\times 10^{13} < & {\w_c^5\varphi_0^4\over ({\rm MeV})^9}Q
<5\times 10^{16}\non\\ 
\textrm{II}: & \iso{3}{18} < & ({m\over\MeV})^{12}Q < \iso{7}{22}\non\\ 
\textrm{III}: & \iso{2}{6} < & |K|({m\over\MeV})^3Q < \iso{2}{7}. 
\eea

\begin{figure}[ht]
\leavevmode
\centering
\vspace*{90mm}
\includegraphics{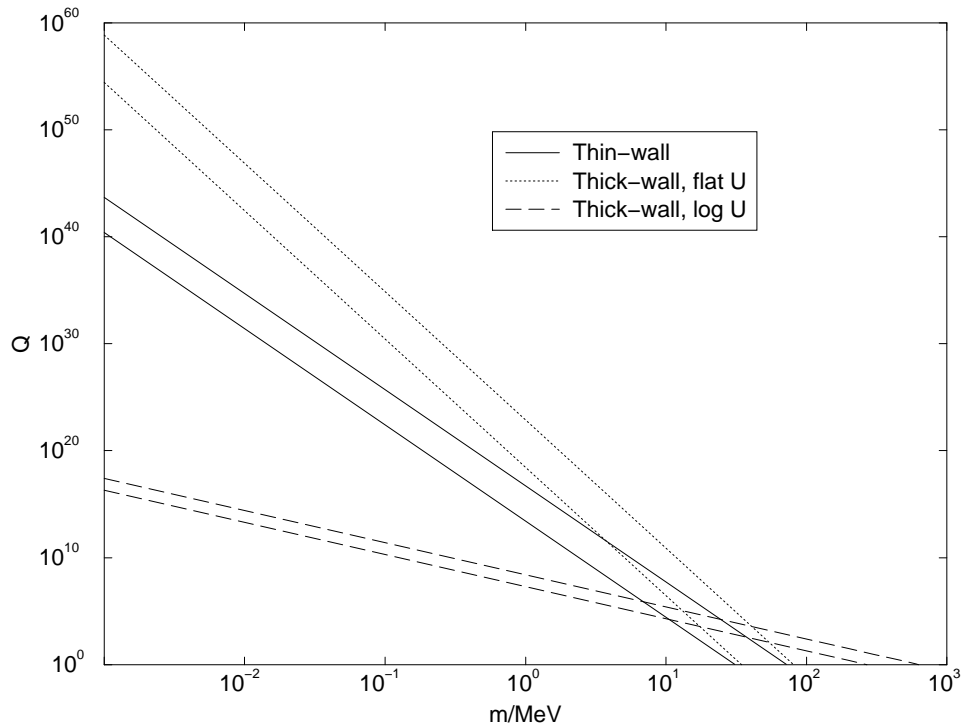}
\caption{Acceptable values (between lines) of $Q$ for different Q-ball types
for  parameter values $\varphi_0=m,\ \w=m,\ K=-0.1$} \label{picqm}
\end{figure}

The allowed range of charge for different types of Q-balls have been plotted
in Fig. \ref{picqm} for different values of the mass parameter $m$.
From the allowed range of charges, we see that the commonly considered
supersymmetric Q-balls that carry baryon number \cite{kusenko418,enqvist538} are unacceptable
as candidates of self-interacting dark matter:
In the gauge mediated case (Type II), $m\sim 10^2-10^4\GeV$, which clearly
leads to unacceptable values of charge.
This is also true for the gravity mediated case (Type III)
with $m\sim 10^2\GeV$, $\w_c\sim m$ and $|K|\sim 0.01-0.1$.
In the thin-walled case (Type I), $\w_c$ is typically of the order
of the supersymmetry breaking scale $\sim 10^2\GeV$ and $\varphi_0$
is at least of the same order (and can be much larger), so that
also in this case the baryon number carrying supersymmetric
Q-balls are not acceptable SIDM candidates.

Q-balls which satisfy the boundaries (\ref{qranges}) can, however, be included
into models where the U(1) -symmetry is not associated to baryon or lepton
number. If the supersymmetry breaking is not setting the scale of parameters
($\omega_c,\ \varphi_0,\ m$), they may be low enough to allow appropriate
values of charge Q. These depend, however, crucially on the details of the
particular model, {\it \e.g.} on the couplings of the Q-balls to the
ordinary matter.

\subsection{Thermally distributed dark matter}
The cross section considered in the previous section do not as such
represent a realistic situation, where one expects a distribution
of Q-balls with different charges. This was considered both
analytically and
in a numerical simulation in \cite{multasimu} in the context
of Type III Q-balls, where it was argued that a Q-ball ensemble
that originates from a fragmentation of a scalar condensate
soon achieves thermal equilibrium. Let us assume that this also
holds true in a general case.
The one-particle partition function reads
\be{z1}
Z_1= \int_{V_D}{d^3xd^3p\over (2\pi)^3}\int_{-\infty}^{\infty}dQ\
\e^{-\beta E+\mu Q},
\ee
where $\mu$ is the chemical potential and $\beta^{-1}=T$ is the temperature
of the Q-ball ensemble, which is related to the average energy
(or average charge) of the system. Assuming that the
distribution is such that the charge both in Q-balls and in anti-Q-balls
is much larger than the net charge of the distribution \ie we set
$Q_+,\ |Q_-|\gg |Q_+ + Q_-|$, we may approximate $\mu\approx 0$, as was the
case in \cite{multasimu}.

Let us write the relation of the mass of the Q-ball to charge as
$M_Q=A Q^B$, where $A$ and $B$ are constants. Then the one particle
partition function reads
\be{z1g}
Z_1={g V_D\over \pi^2B A^{1/B}}\beta^{-3-1/B} G_1(B,0),
\ee
where the function $G_i$ is defined as
\be{gib}
G_i(B,x)=\int_x^{\infty}dt\ t^{i+{1\over B}}K_2(t).
\ee
$K_n(x)$ is the modified Bessel function.
For the Type II $B=3/4$ (see Eq. (\ref{gaugeen})) and for the
Type III $B=1$ (see Eq. (\ref{graveq})), $G_1(B,0)$ can
be evaluated: $G_1(1,0)=3\pi/2$ and $G_1(3/4,0)=(2^{4/3}10/9)\Gamma(2/3)^2$.

We know that the energy density of galactic DM is
\be{rhodm}
\rho_{DM}\approx 0.3 {\GeV\over \cm^3}=2.3\times 10^{-42}\GeV^4.
\ee
The flux of Q-balls is constrained by a number of experiments
probing different values of Q-ball masses,
\be{dmflux}
{1\over 4\pi} n_{DM} v<F_{ex},
\ee
where $n_{DM}$ is the number density of the dark matter constituents,
$v\sim 10^{-3}c$ is their velocity, and $F_{ex}$ is an experimental
constraint. Note that here it is assumed that the velocity distribution
of the dark matter Q-balls in the galactic halo is uniform, only the
mass distribution is assumed to be thermal.

Depending on the details of the Q-balls, some of the Q-balls can be unstable
and hence only a fraction of the total distribution
contributes to the galactic dark matter content.
The charge of the smallest stable Q-ball is denoted here by $Q_{\textrm{\scriptsize stab}}$.
The average mass and number density of stable Q-balls
are easily calculated from Eq. (\ref{z1}).
By approximating $\rho_{DM}\approx n_{stab}\vev{M_{stab}}$ and $\rho_{DM}=
\rho_{stab}$, we can write the constraint Eq. (\ref{dmflux}) as
\be{betalimit}
\beta {G_1(B,\beta A Q_{\textrm{\scriptsize stab}}^B)\over G_2(B,\beta A
Q_{\textrm{\scriptsize stab}}^B)}<{4\pi F_{ex}\over\rho_{DM}v}.
\ee

The expectation value of $s$ can  be evaluated 
%by assuming that $s=CQ^D$, and 
in the limit of small stable Q-balls,
$Q_{\textrm{\scriptsize stab}}=0$ we find
\bea{sexp1}
\vev{s_\ind{I}}&\approx&
0.82({\beta\over\w_c^4\varphi_0^4})^{1/3}~~ {\rm (thin-wall)}\non\\
\vev{s_\ind{II}}&\approx& 1.76 {(\beta m)^{1/3}\over m^3}~~{\rm (thick-wall, flat U)}.
\eea
For  Q-balls in logarithmic potentials, $\vev{s}$ is not calculable in the stable Q-ball limit.
This is due to the fact that the radius of the Q-ball is assumed to constant,
regardless of the charge of the Q-ball, \ie even a zero-charged Q-ball has a
constant radius. This leads to a divergent $s$, which also makes
$\vev{s}$  diverge in the small $Q_{stab}$ limit. Actually $s$ also diverges in the
two other cases as $Q$ tends to zero, but the divergence is milder as can be seen
from Eq. (\ref{diffs}).

To estimate the value of $\vev{s}$ in the $Q_{stab}=0$ limit in the logaritmic case, 
we approximate
\be{logs}
\vev{s}\approx {\sigma_{DD}\over \vev{M_{stab}}}.
\ee

To constrain the flux of Q-balls experimentally we study
the flux of strongly-interacting dark matter particles, which is constrained by several
experiments (see \eg \cite{mcguire}). To study what parameter regions are allowed for
thermally distributed Q-ball dark matter, we adopt a scheme similar to the one
in \cite{kustein}, \ie we assume that the interaction of a Q-ball with 
a nucleon is mediated by a heavy boson Z' so that the interaction cross-section is
given by
\be{sigmaqp}\vspace{0.2cm}
\sigma_{Qp}\sim F ({g\over M_{Z'}})^2 Q^2.
\ee
The form factor $F$ is of order one if the radius of the Q-ball,$R_Q$, is less than
that of the nucleus, $R_n$ and of order $(R_n/R_Q)^3$ if $R_Q\gg R_n$.

\begin{figure}[ht]
\leavevmode
\centering
\vspace*{19cm}
\includegraphics{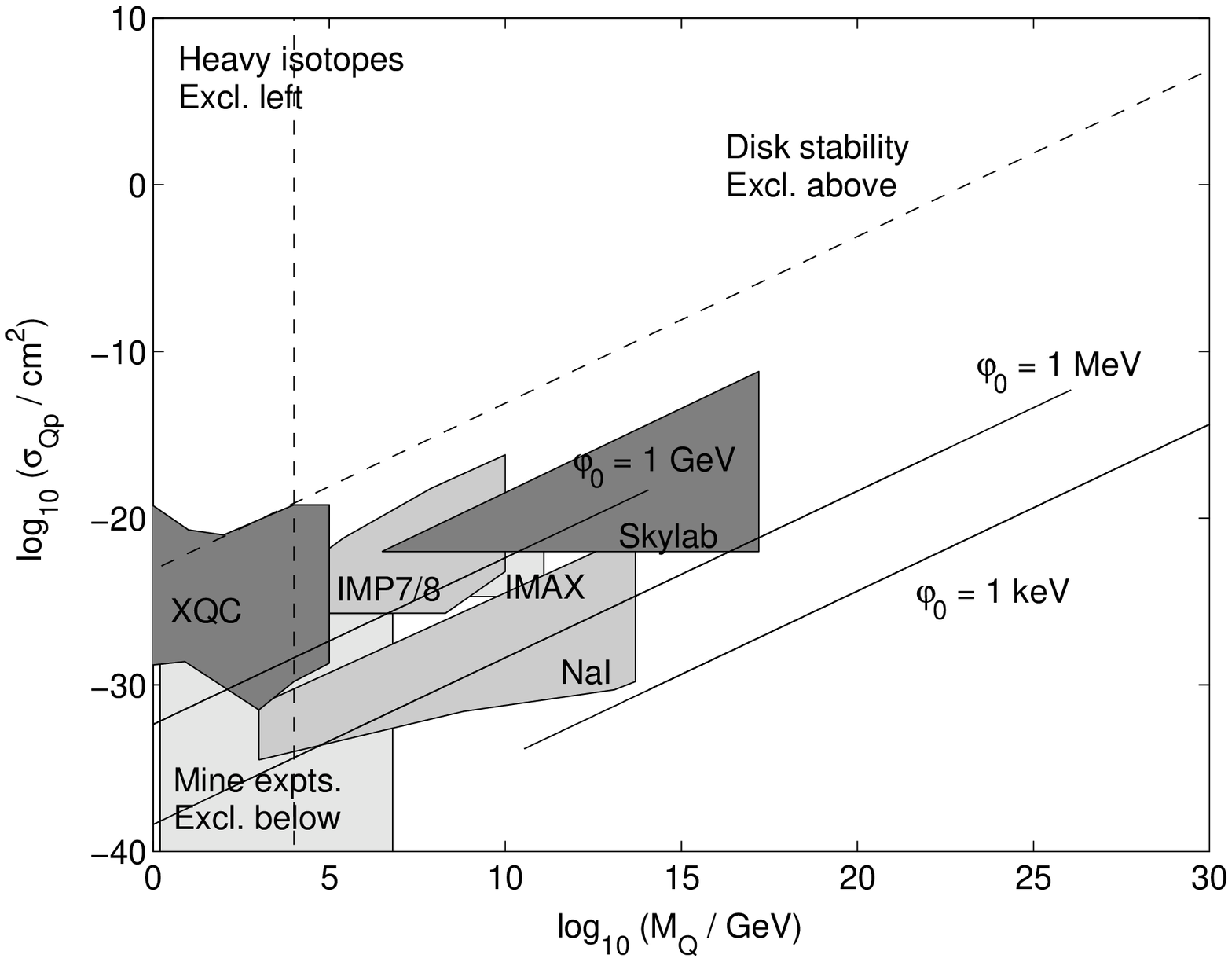}
\includegraphics{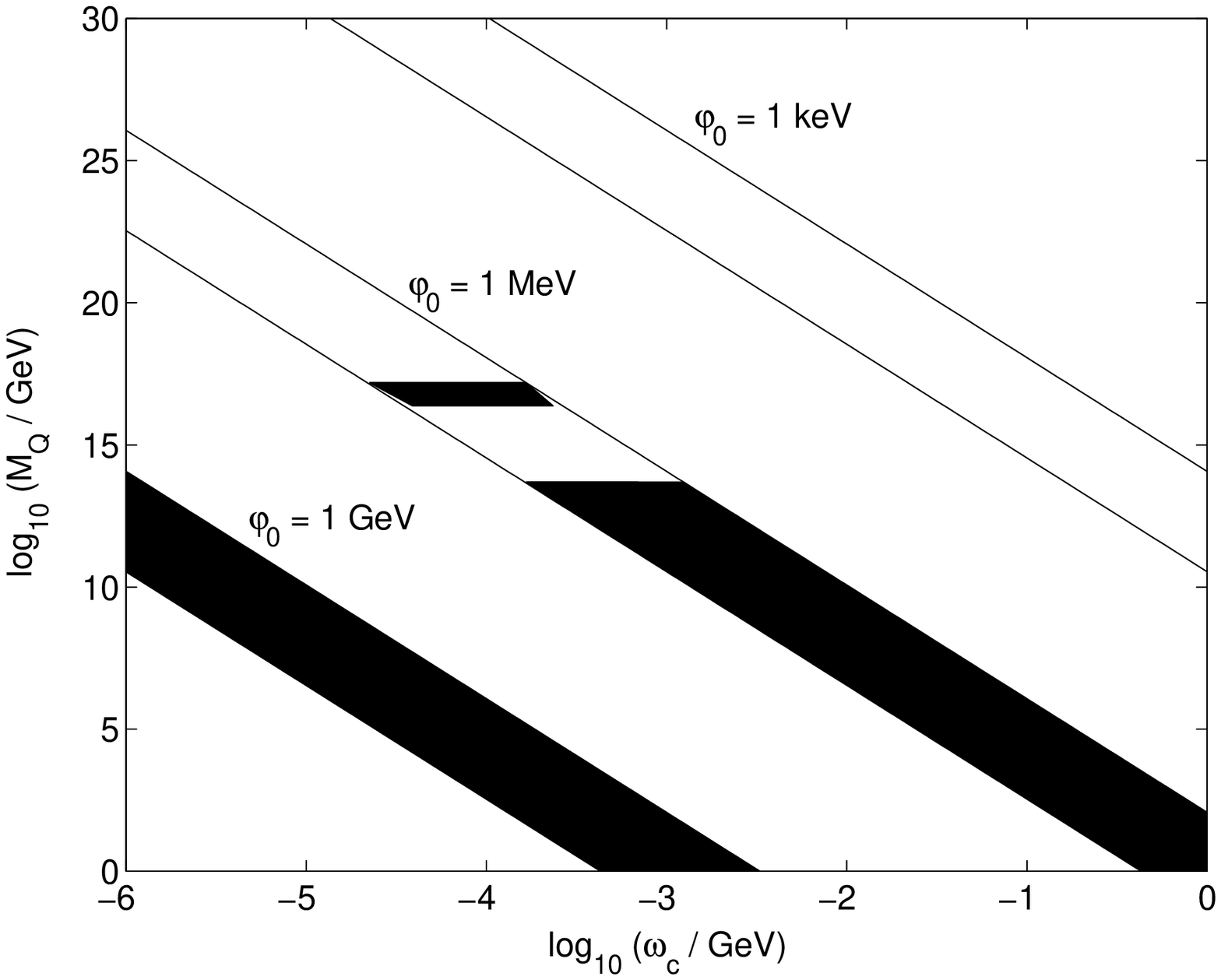}
\includegraphics{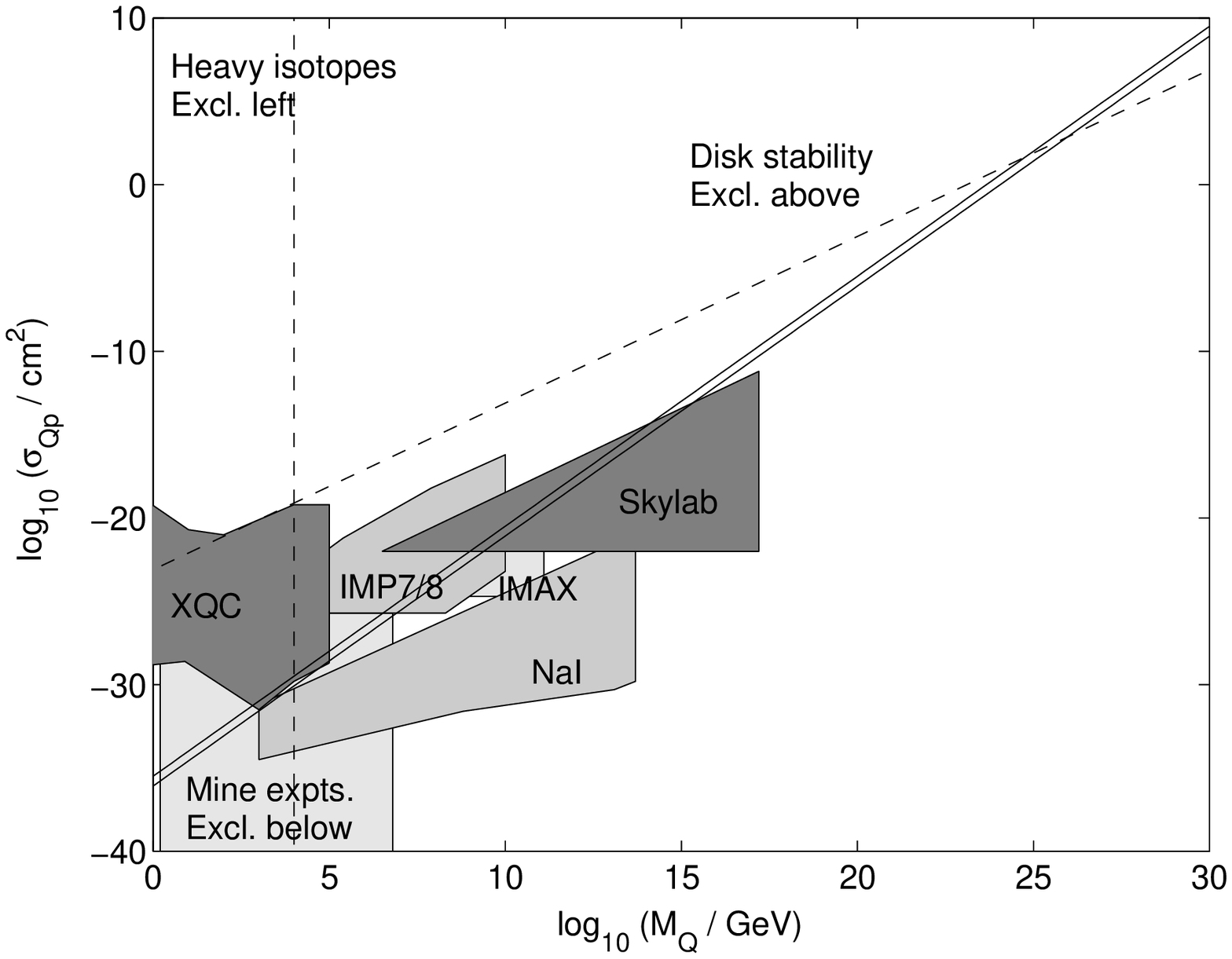}
\includegraphics{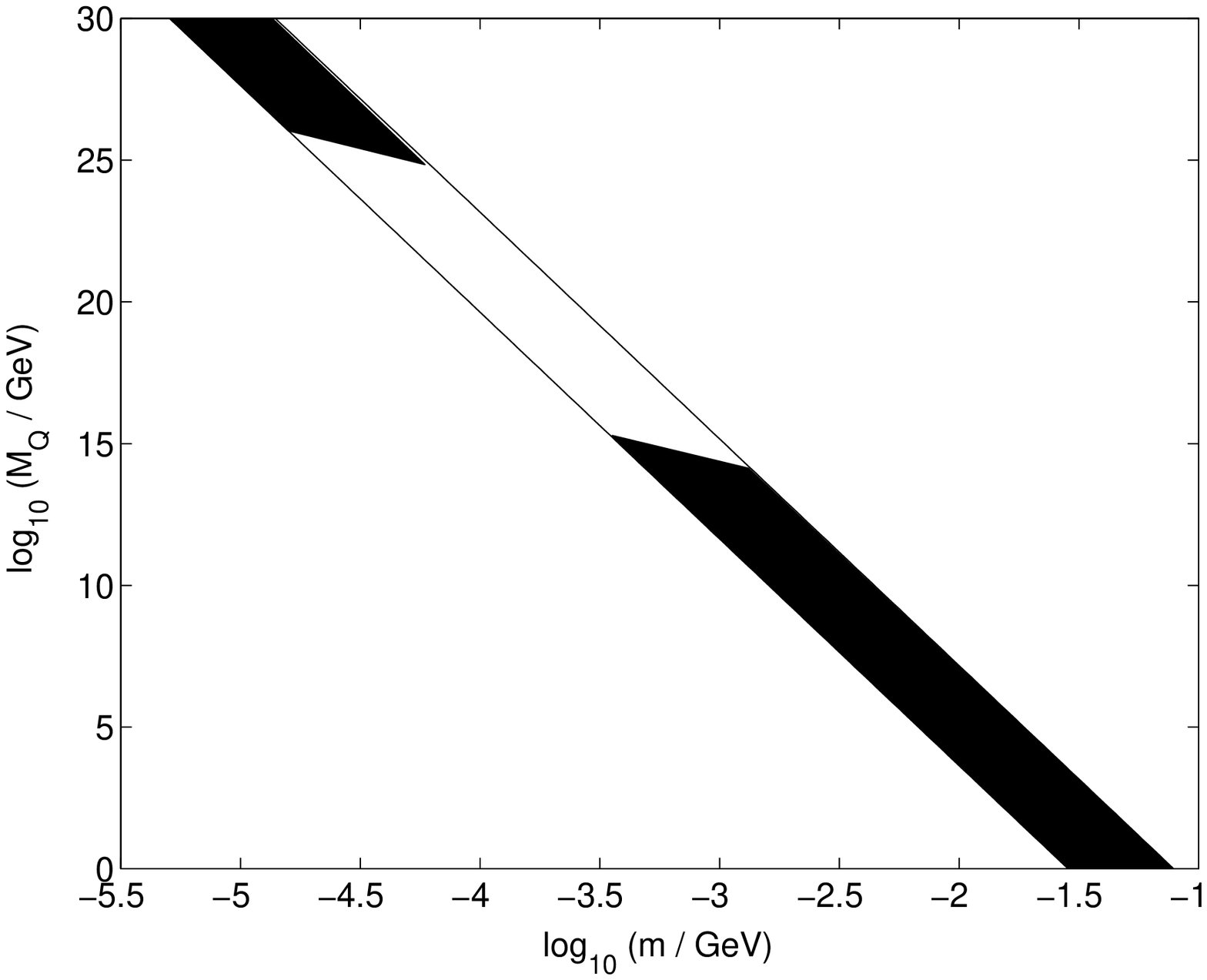}
\includegraphics{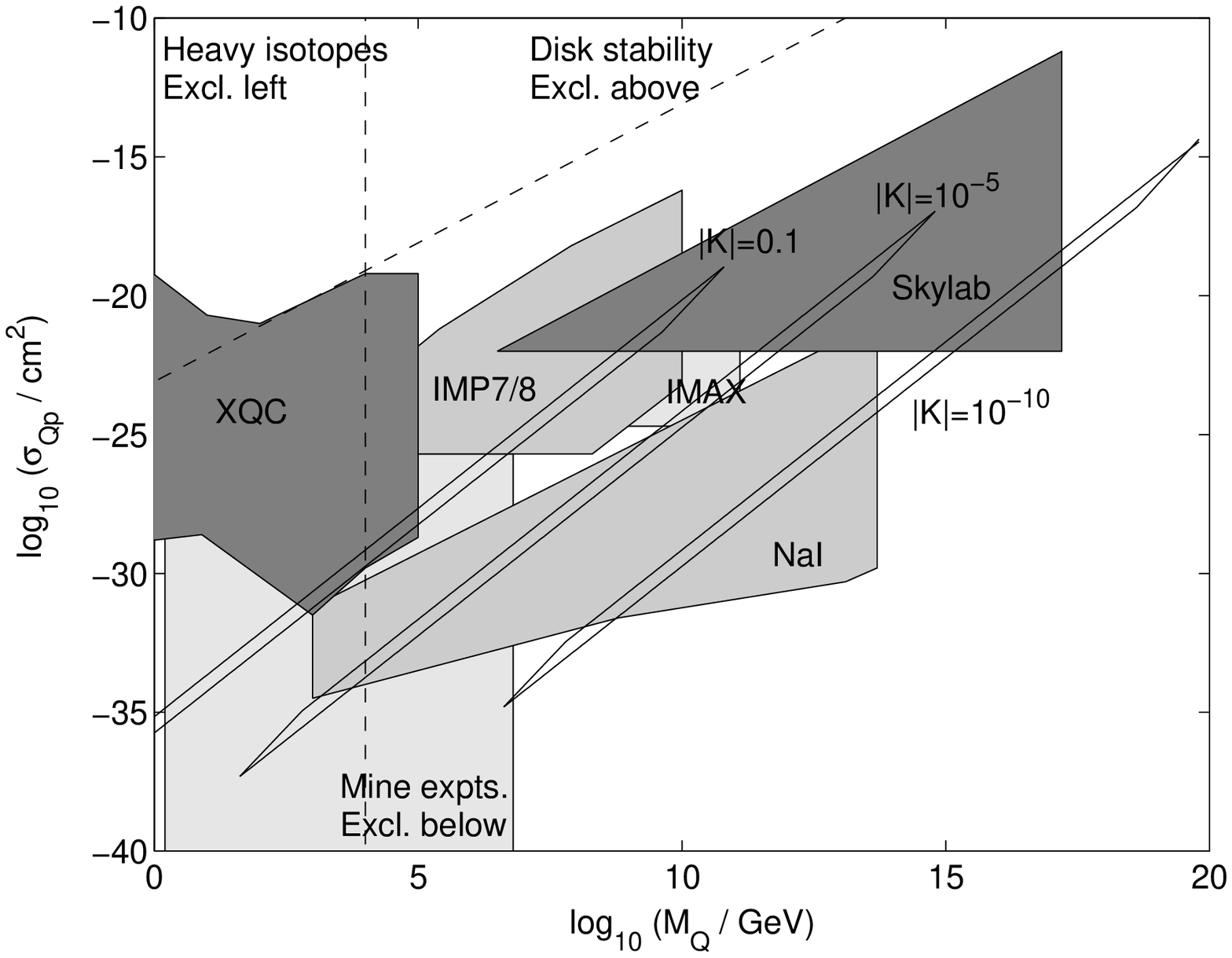}
\includegraphics{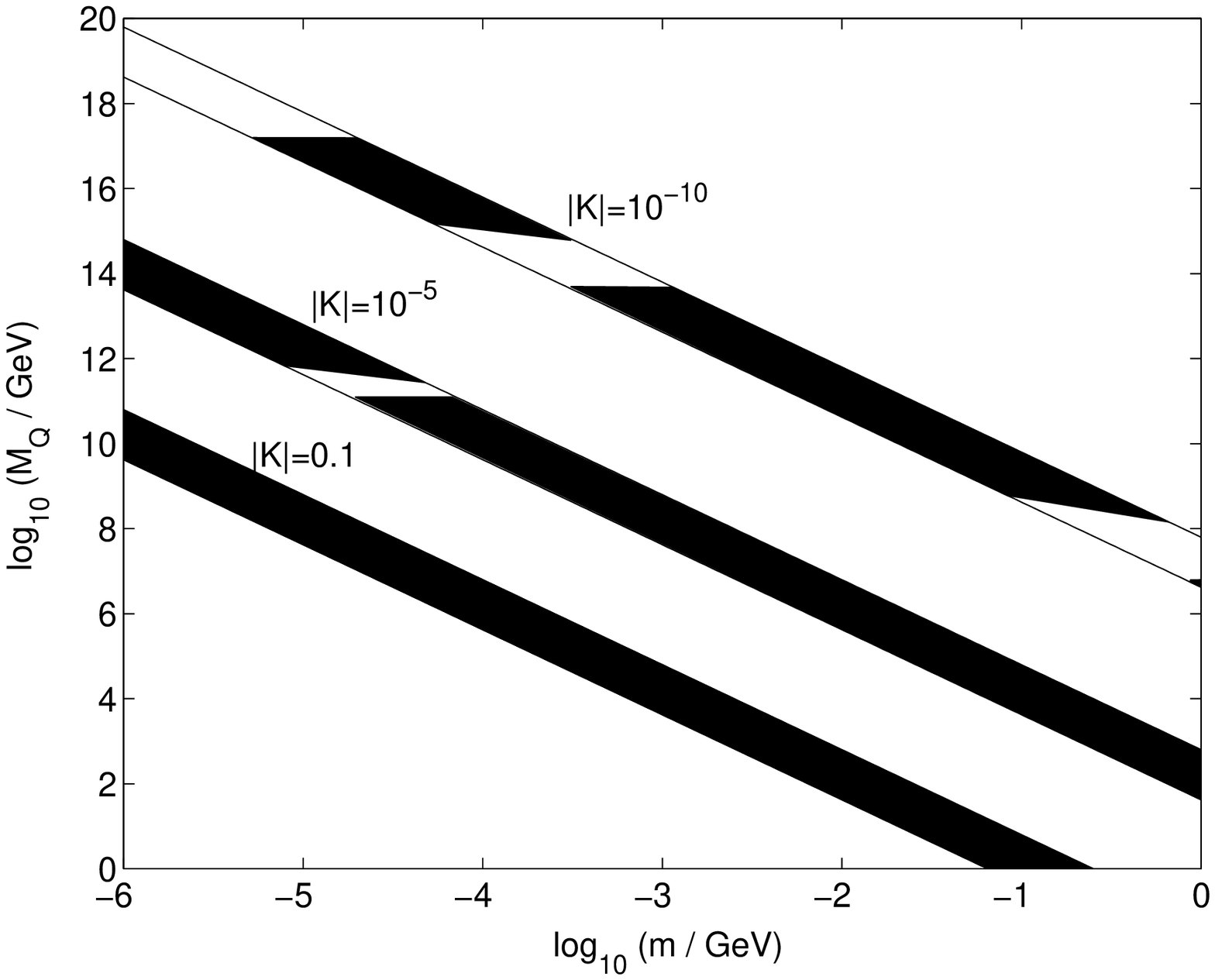}
\caption{Acceptable values for the thin-wall case with $\w_c=1\keV\ldots 1\GeV$ and $\varphi_0=1\keV\ldots 1\GeV$ (the first two figures), flat thick-wall case (the middle figures) and logarithmic thick-wall case (the last two figures) with $m=1\keV\ldots 1\GeV$, $\varphi_0=1\keV\ldots 1\GeV$ and $|K|=0.1,\,10^{-5},\,10^{-10}$. Shaded areas are excluded.} \label{thinw}
\end{figure}

For each Q-ball type we can then calculate the expectation value of 
the mass of the Q-ball, $M_Q$, and of the interaction cross-section, 
$\sigma_{Qp}$. These have been plotted
in Fig. \ref{thinw} for $g=0.1,\ M_{Z'}=1 \TeV$.
The shaded areas are excluded regions in all of the graphs and the allowed
parameter regions are between the solid lines (except in the thin-wall case,
where the allowed region for a fixed $\varphi_0$ is just the line).
In all of the cases in Fig. \ref{thinw}, we have assumed that $Q_{stab}=0$. A
non-zero $Q_{stab}$ shifts curves towards larger $\vev{M_Q}$s as one would expect. 

From Fig. \ref{thinw} it is apparent that the experimentally allowed window
for thermally distributed is narrow in all of the cases. The most promising
case appears to be the thin-walled type Q-ball with a $\varphi_0\gsim\cO(\MeV)$.
The thick-walled Q-balls have less free parameters than in the thin-walled
case and the experiments exclude most of the parameter space. Again, an
appropriate mass parameter is of the order of $\MeV$.

\section{Constraints from Q-ball evolution}
\subsection{Evaporation}

When one considers Q-balls created in the early universe, \eg by the
fragmentation of the AD-condensate \cite{kusenko418,enqvist538}, 
as candidates of SIDM,
the question of Q-ball evaporation as well as thermal effects need
to be addressed.

The evaporation of Q-balls can lead to the washing out of
primordial Q-balls, depending on the details of the theory.
The evaporation rate of a thin-walled Q-ball is bounded from above by \cite{cohen}
\be{evapmax}
{dQ\over dAdt}\leq {\w^3\over 192\pi^2}.
\ee
Evaporation rates for realistic profiles have been considered in
\cite{multaevap}.
The evaporation rate of thin-walled Q-balls is dependent on the combination
$g\varphi_0/\w$, where $g$ represents the coupling between the Q-ball field and
the (massless) fermions that it decays into.
A thin-walled Q-ball with $\w\sim m_\varphi$ then decays at a rate
\be{twdecay}
{dQ\over dt}\approx {m_{\varphi_0}^3\over 48\pi}R^2={1\over 48}
({m_\varphi^7\over\pi^5\varphi^4})^{1/3}Q^{2/3}
\ee
so that its decay time can be roughly estimated by
\be{twdecaytime}
\Delta t\sim {Q\over dQ/dt}=48 ({\pi^5\varphi_0^4\over
m_\varphi^7})^{1/3}Q^{1/3}. \ee

If one takes $m_\varphi\sim \varphi_0$, then $\Delta t\sim
10^2Q^{1/3}/m_\varphi$. Assuming, say, that Q-balls must still be around at
about million years from the Big Bang, $\Delta t_{gf}\sim 10^{37}\GeV^{-1}$.
It is then clear that Q-balls of this type need to be very large not to have
evaporated too early in order to have an effect on the galaxy formation.

In the above calculation it has been assumed, however, that
$g\varphi_0/\w \gsim 1$.
If, on the other hand,
the Q-ball field is very weakly coupled to the fields that it can decay into,
the lifetime of a Q-ball can be long enough for it to play role in
galaxy formation. The small $g$ limit has been studied in \cite{cohen},
where it was found that the evaporation rate tends to
\be{smallgevap}
{dQ\over dtdA}={g\varphi_0\w^2\over 64\pi}
\ee
as $g\rar 0$.
The lifetime of a Q-ball in this limit is of the order
\be{twgtime}
\Delta t\sim {70\over g}({\varphi_0\over m_\varphi^4})^{1/3}Q^{1/3}.
\ee
Hence the coupling constant $g$ needs to small enough to allow long lived
Q-balls. This can be accomplished by fine-tuning or by having the interaction
to be mediated by heavy bosons.

Equations (\ref{evapmax}) and (\ref{smallgevap}) may be used to get
order-of-magnitude approximations for thick-walled Q-balls, too.
Indeed, according to the calculations made in \cite{multaevap} we know that
it gives reasonable order-of-magnitude approximation at least for the
logarithmic potential.

For the thick-wall, flat $U$ Q-ball, applying the bound for the evaporation rate
Eq. (\ref{evapmax}) gives
\be{iirate1}
{dQ\over dt}\leq {\sqrt{2}\pi^2\over 48}m Q^{-1/4},
\ee
\ie large Q-balls evaporate more slowly than small ones.
Assuming that the bound is satisfied for all $Q$ and integrating Eq. (\ref{iirate1})
we get a Q-ball lifetime of
\be{iitime1}
\Delta t={192\over 5\sqrt{2}\pi^2m}Q^{5/4}.
\ee
Again assuming that $\Delta t\sim 10^{37}\GeV^{-1}$, we get a lower bound
on Q-ball charge of $Q\sim 10^{29}(m/\GeV)$. However, if realistic Q-balls
are much smaller, say  $Q \sim 10^{20}$, as simulations suggest 
\cite{kawa2,kawa1,multasimu},  the coupling $g$ should be   $\sim 10^{-22}
{{\rm GeV}\over m}$.

In the weak coupling limit we get
\be{iirate2}
{dQ\over dt}=({\pi\over 8})^{3/2}gmQ^{1/4},
\ee
which translates into a decay time of
\be{iitime2}
\Delta t={4\over 3}({8\over\pi})^{3/2}{1\over gm}Q^{3/4}.
\ee

The radius of the thick-wall, log $U$ Q-ball is independent of charge. Taking
$\w\sim m$, the decay rate is bound by
\be{iiirate1}
{dQ\over dt}\lsim {1\over 48\pi}{m\over |K|},
\ee
and the weak coupling limit gives
\be{iiirate2}
{dQ\over dt}\sim {1\over 16}{g\varphi_0\over |K|}.
\ee
Taking $|K|\sim 0.1$, and assuming that the bound Eq. (\ref{iiirate1})
is satisfied, a Q-ball that survives until $\Delta t_{gf}$ must have
a minimum charge of $Q\sim 10^{36} m/{\rm GeV}$. Again, if
$Q \sim 10^{20}$, the coupling $g$ should be smaller than
$\sim 10^{-26}{{\rm GeV}\over M}$.

From these considerations we can conclude that evaporation can typically
destroy Q-balls before they can affect galaxy formation, unless the
decay channels of the quanta that the Q-balls consist of are greatly suppressed.
As the best candidates then are possibly the Q-balls interacting only
gravitationally, because suppressed interactions arise there naturally.   

\subsection{Thermal Effects}
The thermal bath of the early universe can have an effect on the distribution
of primordial Q-balls by thermally erasing them. Thermal effects on Q-balls
have been considered for different types of Q-balls by utilising various methods
\cite{kusenko418,enqvist538,laine, banerjee, multa511}.
In Refs. \cite{laine, banerjee}
Purely thermodynamical considerations have been utilised to estimate
thermal evaporation rate and diffusion rate.

A different approach has been adopted in \cite{kusenko418, enqvist538, multa511},
where collisions of thermal background particles with a Q-ball have been considered.
In these consideration, two processes are important: dissociation and dissolution.
In dissociation a thermal particle hits a Q-ball and transfers energy to it. If the
rate of energy transfer into the Q-ball is larger than the emission rate of extra energy,
excess energy builds up and can overcome the binding energy of the Q-ball.
In dissolution, a Q-ball loses its charge from the edge of the Q-ball to the
surrounding plasma. A thermal equilibrium exists at the surface of a thick-walled
Q-ball and charge can leave by diffusion.

All of the described thermal processes can in principle alter the initial distribution.
An important factor in the evolution of the Q-ball distribution
is then the reheat temperature, which if large can effectively destroy the
Q-ball distribution. For comparison, for baryonic Q-balls in the gravity-mediated scenario, it has
been estimated that the reheat temperature should be
at most $T_{RH}\lsim Q^{1/4}\ \GeV$
\cite{multa511}. Note, however, that the coupling strentght of the thermal particles to
the Q-ball is crucial: if the coupling is very small, even small Q-balls
can survive the high temperature bath of the early universe.
If Q-balls are to survive evaporation, as discussed above, the coupling
should be weak. Indeed, as was pointed out in Sect. 3.1., SIDM
Q-balls cannot carry have Standard Model interactions (see also \cite{kustein}).
The possible candidates for such Q-ball fields should be searched for either in the hidden
sector, coupled to the Standard Model only via gravity, or in the SM singlet sector of
the extensions of the MSSM.

\section{Conclusions and discussion}

The question of the initial Q-ball distribution is obviously important
when considering the possibility of Q-ball SIDM. Numerical simulations
have shown that if Q-balls form from the fragmentation of an Affleck-Dine
condensate, the initial conditions are important in deciding the characteristics
of the Q-ball distribution \cite{kawa1,kawa2,multasimu}. If the charge and energy
of the condensate are roughly equal, the following distribution consists
only of Q-balls. However, if the condensate carries excess energy compared to the charge,
a large number anti-Q-balls also appear. This has been observed in
simulations by two separate groups \cite{kawa2,multasimu}.
Appearing anti-Q-balls obviously lead one to consider the possibility of 
dark matter annihilations and its effect on galaxy formation.

We should also mention that
there exist other suggestions to alleviate the problems of standard
CDM models. Among these are the ideas of decaying dark matter \cite{cen} and
annihilating dark matter \cite{craig}. In the decaying dark matter model
some of the dark matter particles decay into relativistic particles by $z=0$
so that small dwarf galaxies fail to form.
In \cite{craig} the evolution of an isolated dark matter halo which
undergoes both scattering and annihilation was considered.

Q-balls can exhibit all of the three phenomena: evaporation, annihilation and scattering,
and hence one can speculate that dark matter consisting of interacting Q-balls
can exhibit different types of behaviour. How to actually combine all of the
elements of Q-ball dynamics so that appropriate galactic halos are produced is
a question that clearly requires further study.

Thus, in an attempt to develop a scenario where primordial Q-balls act as
self-interacting dark matter, one must then pay attention to several issues.
First of all, an efficient mechanism of producing appropriate Q-balls needs
to exist. The fragmentation of the AD-condensate is a promising candidate for
such a mechanism. Secondly, the produced Q-balls must survive thermal effects
if they are to influence galaxy formation. On the other hand, thermal effects
can also be responsible for erasing unwanted small Q-balls.
One must also keep in mind evaporation processes which can lead to an
early decay of Q-balls. This might also offer an exciting scenario where
Q-balls that have played a significant role in galaxy formation, have since
decayed leaving particle dark matter in their place.
The obvious question that requires an answer is the composition and interactions
of the dark matter Q-balls. The commonly studied baryonic Q-balls are not
acceptable so one must develop a model where the requirements for cross-section
and mass are acceptable, while keeping in mind the experimental constraints.

On the basis of the results of this paper, the natural scale of the Q-ball scalar
particle appears to be $\MeV$, regardless of the type of the Q-ball
considered. The charge of an appropriate dark matter Q-ball depends on the
considered Q-ball type and very strongly on the parameter values, as Fig.
\ref{picqm} shows. In each case, the mass parameter must be close to $\MeV$,
otherwise charge can become unacceptably small or large. On the other hand,
the charge of a dark matter Q-ball can then vary greatly, which obviously has
an effect on the possibility of its detection. In \cite{kustein} it was
suggested, by applying naturalness arguments, that the charge of a dark matter
Q-ball is of the order of $10-10^3$ in the thin-wall case and $10^4-10^5$ in the
flat potential case. It should be noted that the charge can be much greater
and considerations should not be limited to small values of charge. Furthermore,
naturalness arguments which indicate $\w_c\sim\varphi_0$, need to be critically
considered in the thin-wall case, where $\w$ can differ greatly from $\varphi_0$.

In \cite{kustein} a uniform Q-ball distribution was discussed.
If Q-balls form from the fragmentation of the the Affleck-Dine condensate,
the initial conditions of the condensate are decisive in determining whether
a uniform charge distribution is a good approximation or not. In this paper
we have also considered the possibility of a thermal Q-ball charge distribution,
which, on the basis of the simulations, is a realistic possibility.
The appropriate Q-ball mass parameter obviously depends on the experimental
flux limit, but is again naturally of the order of $\MeV$ or slightly less.
Such a small mass scale might be difficult to achieve naturally in the
extensions of MSSM but is not necessarily a problem for hidden sector
Q-balls.

The experimental flux limit is dependent on the composition of the Q-balls
and their interactions with matter. If, for example, Q-balls reside in the
hidden sector and interact only gravitationally, their flux can be very high.
If, however, they do have also other interactions with ordinary matter,
experimental flux limits can be calculated. In the case of thermally
distributed Q-ball dark matter, the parameter space appears to be
constrained, especially for thick-walled Q-balls.

As it was discussed, evaporation can be a decisive process in determining
whether Q-balls can act as dark matter or not. The exact charge limits
coming from evaporation processes obviously again depend on the details,
but it seems that if a decay channel to a light particle exists,
Q-balls must be large or their couplings extremely suppressed in order to act as
dark matter during galaxy formation. Even then, it seems probable that they
would have decayed by now, leaving particle dark matter in their place.

%%%%%%%%%%%%%%%%%%%%%%%%%%%%%%%%%%%%%%%%%%%%%%%%%%%%%%%%%%%%%%%%%

\section*{Acknowledgements}
This work has been supported by the Academy of Finland under the contract 101-35224 and by the Finnish Graduate School in Nuclear and Particle Physics.

%%%%%%%%%%%%%%%%%%%%%%%%%%%%%%%%%%%%%%%%%%%%%%%%%%%%%%%%%%

\end{document}